\pdfoutput=1
\documentclass[aps,pre, twocolumn, groupedaddress]{revtex4-1}

\usepackage{lipsum}
\usepackage{float}
\usepackage{mathtools}
\usepackage{graphicx}
\usepackage{subfigure}
\usepackage{caption}
\usepackage{dcolumn}
\usepackage{amsmath}    
\usepackage{amssymb}
\usepackage{bm}
\usepackage{breqn}
\usepackage{hyperref}
\usepackage{latexsym}
\usepackage{verbatim}
\usepackage[normalem]{ulem}
\usepackage{color}
\usepackage{parskip}
\def\beq{\begin{equation}}
\def\eeq{\end{equation}}

\def\beq{\begin{equation}}                          
\def\eeq{\end{equation}}                          
\def\bea{\begin{eqnarray}}                          
\def\eea{\end{eqnarray}}

\DeclareRobustCommand{\uvec}[1]{{%
  \ifcsname uvec#1\endcsname
     \csname uvec#1\endcsname
   \else
    \bm{\hat{\mathbf{#1}}}%
   \fi
}}
\preprint{}
\begin{document}
\preprint{}

\title{Kinetics and steady state of polar flock with birth and death}
\author{Pratikshya Jena$^{1}$}
\email{pratikshyajena.rs.phy20@itbhu.ac.in}
\author{Shradha Mishra$^{1}$}
\email{smishra.phy@iitbhu.ac.in}
\affiliation{ Indian Institute of Technology (BHU)}
\date{\today}

\begin{abstract}
    {We study a collection of polar self-propelled particles or polar flock  on a two dimensional substrate with birth and death. 
    Most of the previous studies of polar flock with birth and death have assumed the compressible flock, such that the local density of flock is completely ignored. Effect of birth and death
    of particles on the flock with moderate density is focus of our study.  System is modeled using coarse-grained hydrodynamic equations of motion for local density and velocity of the flock and solved using numerical integration of the nonlinear coupled partial differential equations of motion and linearised hydrodynamics about the broken symmetry state. We studied the ordering kinetics as well as the 
    steady state properties of the immortal flock and flock with finite birth and death rate. The ordering kinetics of the velocity field remains unaffected whereas the density field shows a crossover from asymptotic growth exponent $5/6$ for the immortal flock to diffusive limit $1/3$ for large birth and death rates. In the steady state, the presence of birth and death rate leads 
    to the suppression of speed of sound wave and density fluctuations in the system. }
    \end{abstract} 
 \maketitle
\section{Introduction \label{Introduction}}
Active mater \cite{ramaswamy2010mechanics, marchetti2013hydrodynamics, toner2005hydrodynamics, bechinger2016active, vicsek1995novel,semwal2021dynamics} is a novel class of nonequilibrium systems which consist of interacting units that consume energy individually and generate motion and mechanical stress collectively. Each individual unit contains  an intrinsic asymmetry and  have tendency to  self-propelled along the direction of asymmetry. Such systems span a huge range of length scales, starting from very small scales of few microns sized molecular motors, bacterial colonies, cells, cytoskeleton filaments  etc. to large scales of the order of few meters and kilometers such as bird flock, human crowd, animal herd etc. \cite{cichos2020machine, needleman2017active, de2015introduction, fodor2018statistical, balasubramaniam2022active, feder2007statistical, bottinelli2017jammed}. The collection of such particles show interesting features such as pattern formation \cite{liebchen2017collective}, nonequilibrium disorder to order transitions \cite{mahault2018self, bhattacherjee2015topological, chate2008collective, mishra2017boundary, singh2021ordering,dikshit2022activity}, anomalous fluctuations \cite{liverpool2003anomalous} etc., those are not present in corresponding analogous equilibrium systems.\\ 
Flocking \cite{toner2005hydrodynamics} or collective and  coherent motion of large number of organisms is the most interesting and  ubiquitous phenomenon in many biological systems. 
Unlike the corresponding equilibrium system \cite{mermin1966absence}, flock shows a true  long ranged, long time behavior in two dimensions. The appearance of dense ordered clusters of particles, propagating  sound modes and the giant number fluctuation are observed \cite{tu1998sound} in flock in the steady state. \\
Most of the previous studies of polar flock focused on the immortal flock, where number of particles remain constant \cite{toner1995long, toner1998flocks, chate2008collective, vicsek1995neovl}. In a recent study \cite{toner2012birth}, the author studied a system of polar flock without number conservation and showed that the system behaves differently from the immortal flock i.e, the longitudinal velocity fluctuations in the direction of flock motion replace the sound modes of number conserving flock. The study of \cite{toner2012birth} focused on the high density limit of polar flock with birth and death, such that the density field is completely ignored. The properties of immortal flock with finite density is not explored, and it is the focus of our study.\\
In this letter, we study the polar flock  without number conservation (Malthusian flock) \cite{malthus1992malthus}. We model the collection of polar self-propelled particles with birth and death. We write the coarse-grained hydrodynamic equations of motion for local density and velocity and study it using numerical solution of nonlinear coupled partial differential equations of motion and linearised hyrodynamics about the broken symmetry state.\\ 
Starting from random homogeneous ordered state of density and velocity, we study the ordering kinetics of both the fields and find that the ordering of velocity field remains unaffected due to birth and death, whereas the density field shows a strong dependence on the birth and death rate. Density shows an asymptotic growth law of $5/6$ \cite{toner2012birth} for immortal flock and low birth and death rate. For large birth and death rate the approach to asymptotic growth is faster and towards the diffusive growth law $1/3$ \cite{bray2002theory}. In the steady state, the presence of birth and death rate suppresses the speed of traversing sound mode and density fluctuations in the system.\\
   \begin{figure*}
  \centering
  \includegraphics[scale=0.5]{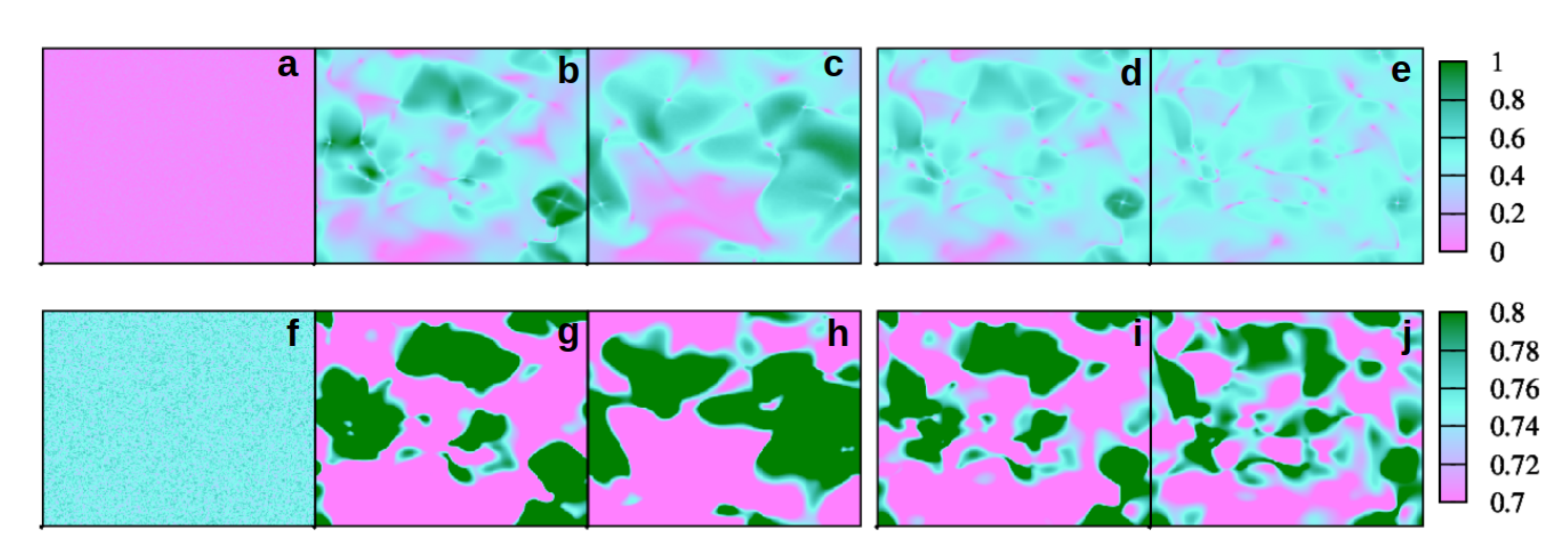}
  \caption{(color online) Real space snapshots of local velocity $v({\bf r}, t)$ (upper panel) $(a-c)$ and density $\rho({\bf r}, t)$ (lower panel) $(f-h)$ for the IF and at times $t=0, 900$ and $200$ respectively. $(d-e)$ and $(i-j)$ are the snapshots for local velocity  and density respectively at fixed time $t=900$ and for different $\lambda = 0.05$, $0.1$.}
  \label{fig:fig1}
  \end{figure*}
  \section{Model  \label{Model}}
We model a collection of polar self-propelled particles with birth and death. We approach the problem using
coarse-grained hydrodynamic equations of motion for two slow variables, viz .(i) local density $\rho({\bf r}, t)$ and (ii) velocity ${\it {\bf v}}({\bf r}, t)$ of particles at position ${\bf r}$ and at time $t$. The hydrodynamic equation
for the density $\rho({\bf r}, t)$ is:
\begin{equation}
\partial{_t}\rho+ v_0 \nabla \cdot ({\it {\bf v}} \rho)=D_{\rho} \nabla^2 \rho + g(\rho)
\label{eq:eqrho}
\end{equation}

and for the local velocity ${\it {\bf v}}({\bf r}, t)$ is:
   \begin{equation}
  \begin{aligned}
   \partial{_t}{\bf v}+ {} &\lambda{_1}({\it {\bf v}}. \nabla){\it {\bf v}}+\lambda{_2}(\nabla \cdot {\it {\bf v}}){\it {\bf v}} =\ \alpha(\rho) {\it {\bf v}} \\ & \ -\beta  \mid{\it {\bf v}}\mid^{2}{\it {\bf v}}-\sigma_{1}\nabla \rho +D{_T}\nabla^{2}{\it {\bf v}}+{\it {\bf f}} 
\end{aligned}
\label{eq:eqv}
\end{equation}

The density equation is continuity equation with addition to birth and death. The term on the left hand side is the current due to self-propulsion. The first term on the right side is the diffusion current  and the last term is an additional birth and death term of the 
form $g(\rho)$=$\lambda \rho(1-\frac{\rho}{\rho_{0}})$. 
The parameter $\lambda$ controls the rate of birth and death. For $\lambda=0$, the model
reduces to the immortal flock (IF) \cite{toner1998flocks} and for finite 
$\lambda$ we have flock with birth and death (MF) \cite{malthus1996essay,doi:10.1063/5.0086952}. Here, $\rho_{0}$ is constant that controls the mean density.
The velocity Eq.\ref{eq:eqv} is the same as introduced by Toner and Tu for the immortal flock \cite{toner1998flocks}. 
In Eq.\ref{eq:eqv}, the right hand side, the terms, $\alpha(\rho) \bf{v}$, $\beta  \mid{\it {\bf v}}\mid^{2}{\it {\bf v}}$ and $D_{T}\nabla^{2}{\it {\bf v}}$ arise in analogy with the  time dependent Ginzburg-Landau equation (TDGL) \cite{goldenfeld2018lectures} which is responsible for ordering towards the broken symmetry phase.
$\lambda_{1}( {\bf v} \cdot {\nabla}) {\bf v}$, $\sigma_{1}\nabla \rho$ terms are analogous to the convection and pressure due to gradient in density respectively and similar to the terms present  in Navier-Stoke's equation for fluid dynamics. $\lambda_{2}(\nabla \cdot {\bf v}){\bf v}$ is another convective non linearity appears due to absence of Galilean invariance in the system.  
 The last term in Eq.\ref{eq:eqv} ${\bf f}({\bf r},t) = (f_1, f_2)$  is a random driving force representing noise and it is Gaussian with white-noise correlation and strength $\Delta$;
\begin{equation}
    \langle f{_i}({\bf r},t)f{_j}({\bf r^{'}},t{'}) \rangle = \Delta\delta_{ij}\delta({\bf r}-{\bf r}^{'})\delta(t-t^{'})
\end{equation}
Numerical integration of Eq.\ref{eq:eqrho} and \ref{eq:eqv} is performed with  homogeneous initial density (with mean  $\rho_0 = 0.75$) and random velocity in a box of size $K \times K$ with periodic boundary condition in the both directions.
The integration is performed using Euler's scheme \cite{bally1996law} with $\Delta x = 1.0$ and $\Delta t =0.1$.
  The other parameters of the system are fixed to $\beta=1.0$, $\lambda_{1}=0.5$, $\lambda_{2}=1.0$, $\alpha(\rho)=(1-\rho/\rho_c)$ ($\rho_c=0.5$), $D_{\rho}=1.0$, $\sigma_{1}=v_{0}/\rho_0$, $D_{T}=1.0$, $v_0=0.5$ and varied the birth and death rate   $\lambda$  from $0$ to $1.0$. We checked that our numerical integration is stable for the above choice of parameters.
The simulation is carried out by varying the system size i.e $K=256$ to $2048$. Starting with random homogeneous state, system is quenched to the above parameters such that steady state  is an ordered state. We studied the ordering kinetics of the  velocity and the density field and properties of the steady state for different birth and death 
rate $\lambda$.
 Steady state results are obtained for $K=256$ to  $1000$  for time $t=2 \times 10^5$ whereas coarsening studies are performed on 
 relatively large system size $K=2048$ with final time $t=7 \times 10^3$. For better statistical averaging results are averaged over $100$ independent
 realisations.
\section{Result\label{Result}}
We first plot the time evolution of real space snapshots of the system, initiated with random homogeneous state. Fig.\ref{fig:fig1} show the time evolution of local velocity
${v({\bf r}, t)} = |{\bf v}|$ (upper panel) and magnitude of local density ${\rho({\bf r}, t)}$ (lower panel). Fig.\ref{fig:fig1}($a-c$; $f-h$) shows the plot  for the immortal flock at three different times $t=0$, $900$ and $2000$ and Fig \ref{fig:fig1}($d-e$; $i-j$) is for the fixed time $t=900$ and different $\lambda$=, $0.05$ and $0.1$ respectively. As observed from the Fig.\ref{fig:fig1}($a-c$), the velocity field  shows that system evolves towards the ordered state with the growing domains of ordered region with large magnitude of local $v$. Also the domains of high density region grows with time as shown in Fig.\ref{fig:fig1}($f-h$). The growing domains of ordered velocity and high density is also affected by increasing $\lambda$ as shown in Fig. \ref{fig:fig1}($d-e$; $i-j$) for fixed time and varying $\lambda$. The growth is more affected for density Fig. \ref{fig:fig1}($i-j$) in comparison to velocity field Fig. \ref{fig:fig1}($d-e$).\\
The characteristics of the growing domains  is measured by  calculating the two-point velocity and density correlation functions $C_{v}(r, t)=\langle {\bf v}({\bf r}_{0}, t) \cdot {\bf v}({\bf r}_{0}+{\bf r}, t) \rangle$ and $C_{\rho}(r, t)=\langle \delta\rho({\bf r}_{0}, t) \delta\rho({\bf r}_{0}+{\bf r}, t) \rangle$ respectively. 
Here, $\delta \rho$ represents the fluctuation in density from the mean $\rho_0$ and the $<..>$ means averaging over directions, reference points ${\bf r}_0$ and over $100$ independent realisations. 
      \begin{figure}
    \centering
    \includegraphics[scale=0.30]{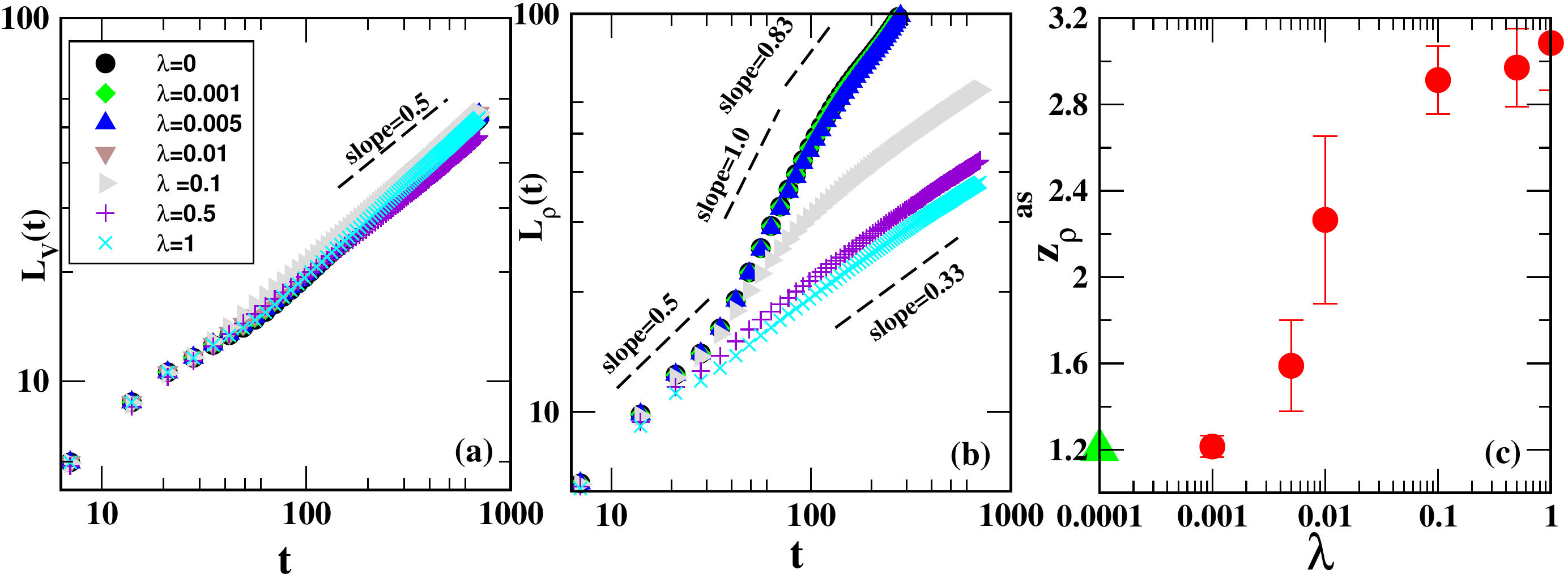}
      \caption{(color online) (a,b) represent the $\log-\log$ plot of characteristic lengths $L_{v, \rho}(t)$ {\em vs.} time $t$ for velocity and density field respectively for different $\lambda$ values. The dashed lines have their respective slopes as shown on top of them.
      (c) plot of $z_{\rho}^{as}$ {\em vs.} $\lambda$. The solid $\triangle$ on the $y-$axis is value $1.2$.} 
    \label{fig:fig2}
    \end{figure}
  \begin{figure}
    \centering
    \includegraphics[scale=0.35]{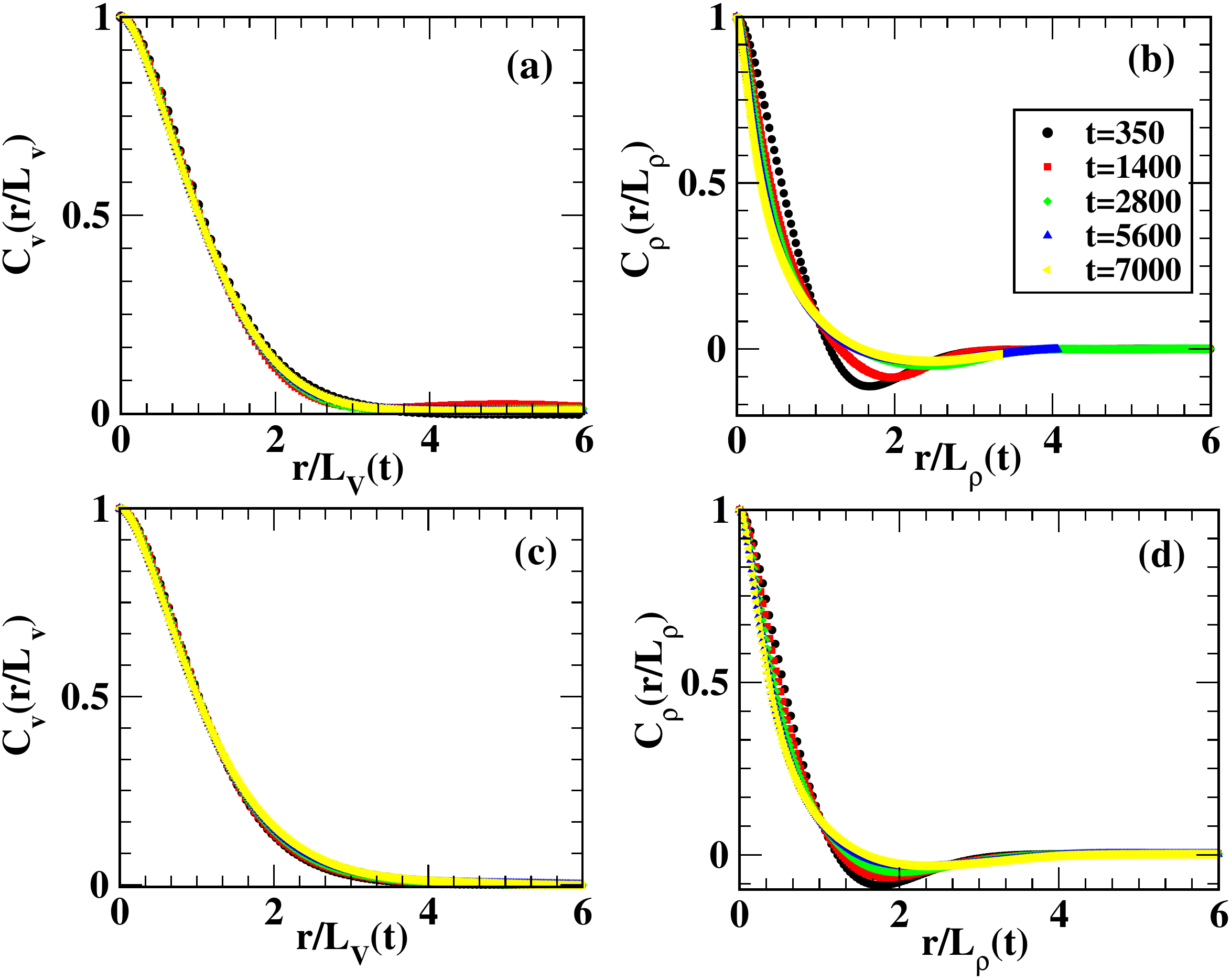}
     \caption{(color online) The plot of scaled two-point correlation functions $C_{v, \rho}(r/L_{v, \rho}(t))$ {\em vs.} scaled distance $r/L_{v, \rho}(t)$ for velocity  and density fields.  $(a-b)$ is for the IF ($\lambda=0$) and $(c-d)$ is for the MF with $\lambda=0.01$. Different colors are for different times as shown in the legend.}
     \label{fig:fig3}
    \end{figure}
  \begin{figure}
    \centering
    \includegraphics[scale=0.30]{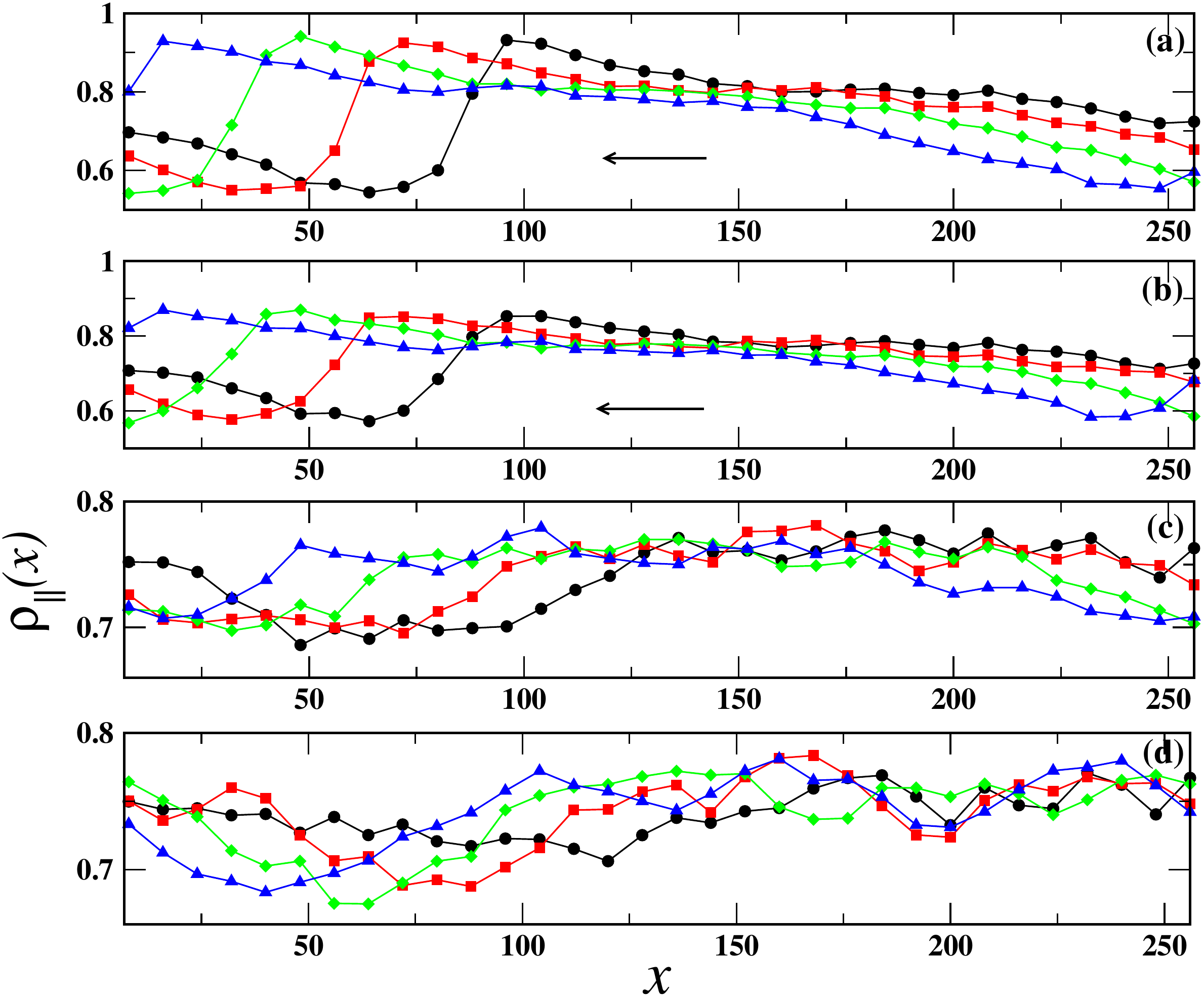}
    \caption{(color online) $(a-d)$, the plot of $\rho_{||}(x)$ {\em vs.} $x$ for $\lambda = 0$,  $0.0005$, $0.0008$, $0.001$  respectively. Different colors/symbols  $\circ$ (black),  $\square$  (red), $\diamond$ (green) and $\triangle$ (blue) in each window represent the different times ($t_1 < t_2 < t_3 < t_4$) respectively in the steady state at the interval of $t=400$. 
    The arrows in $(a-b)$ show the direction of moving flock. }
    \label{fig:fig4}
    \end{figure}
           \begin{figure} 
  \centering
  \includegraphics[scale=0.31]{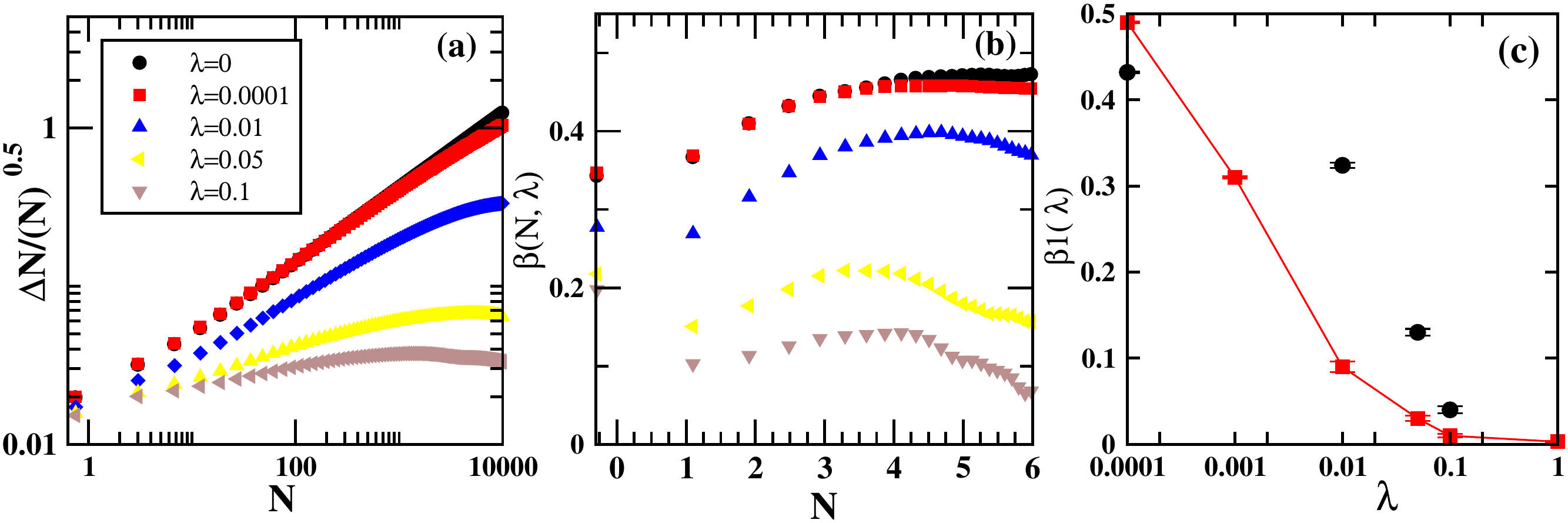}
 \caption{(color online) (a) The $\log-\log$ plot of $\frac{\Delta N}{\sqrt{N}}$ {\em vs.} $N$ for different $\lambda$. (b) The plot of $\beta(N, \lambda)$ {\em vs.} $N$.
 (c) The plot of $\beta(\lambda)$ {\em vs.} $\lambda$.
The black data are obtained from numerical simulation and data with line (red)  obtained from Eq. 21  in \ref{appendix} by putting all the constants from the numerical simulation.} 
  \label{fig:fig5}
  \end{figure} 
{\em Ordering kinetics}:- 
We define  the size of the growing domain of by calculating the characteristics lengths $L_{v, \rho}(t)$  
defined  by the distance by which the  the two-point 
correlation functions (as shown in Fig.1(a-b) (SM1)) decay to $0.1$ of its value at $r=0$. Fig.\ref{fig:fig2}(a-b) respectively  shows the plot of the two characteristic lengths $L_{v,\rho}(t)$ {\em vs.} time $t$ for different $\lambda$'s from $0$ to $1$. We find that the growth of $L_v(t)$ remains almost independent of $\lambda$. For all the cases the 
asymptotic growth law is $0.5$ as shown by the dashed line. Hence the kinetics of velocity remains unaffected due to the birth and death parameter $\lambda$ and remains the same as for the usual non-conserved growth kinetics of  {\em Model A} \cite{hohenberg1977theory, das2018ordering}.  As shown in Fig.\ref{fig:fig2}(b) the growth of $L_{\rho}(t)$ shows strong dependence on the parameter $\lambda$. For immortal flock and  weak birth and death $\lambda  \le 0.01$, the growth of $L_{\rho}(t)$ clearly shows three regimes. For very early time, the growth is mostly dominated by  background non-conserved continuous velocity field  and $L_{\rho}(t) \simeq t^{0.5}$, as we increase time, density starts to grow much faster and $L_{\rho}(t) \simeq t$. The fast growth is due to the hydrodynamic modes present in moving ordered flock \cite{toner1995long, toner2005hydrodynamics,chate2008collective}. Then at the late times 
$L_{\rho}(t)$ approaches to the asymptotic growth law $\simeq t^{5/6}$ predicted for the continuum theory  of polar  flock  \cite{toner1998flocks}. On increasing $\lambda>0.01$, the approach to the asymptotic growth happens early and asymptotic growth exponent decreases with increasing $\lambda$ and finally for large $\lambda$'s it approaches to the diffusive limit  with $L_{\rho}(t) \sim t^{1/3}$ for the conserved growth kinetics \cite {lifshitz1961kinetics}.  In Fig.\ref{fig:fig2}(c) we show the plot of asymptotic growth exponent $z_{\rho}^{as}$ {\em vs. }$\lambda$, where $z_{\rho}^{as}$ is obtained from the late time power law dependence of $L_{\rho}(t) \simeq t^{1/z_{\rho}^{as}}$ in the main figure. We found that for very small $\lambda <0.001$ , $z_{\rho}^{as} \simeq 6/5$ and then it increases with increasing $\lambda$ and saturate to the diffusion limit $\simeq 3$ for higher $\lambda$. Hence asymptotic growth of the density is driven by the conventional sound mode propagating in the direction parallel to the direction of flock; for the IF and flock with small birth and death rate, whereas it is purely diffusive for the MF.\\
We further characterise the scaling behaviour of two-point correlation 
functions. In Fig.\ref{fig:fig3}(a-b) we show the plot of scaled correlations for velocity and density $C_{v}(r/L_{v}(t))$ and $C_{\rho}(r/L_{\rho}(t)$ respectively for the  IF, $\lambda=0.0$  and in Fig.\ref{fig:fig3}$(c-d)$ for the $\lambda=0.01$ and for different times. We find that for both the cases  IF  and $\lambda=0.01$, the  velocity field $C_{v}(r/L_{v}(t))$ shows the good scaling collapse, whereas 
density $C_{\rho}(r/L_{\rho}(t))$ does not. Non scaling of density 
field for immortal flock is also reported in previous studies \cite{das2018ordering}, is due to the presence of more than one regime of growing domain as shown in $L_{\rho}(t)$ in Fig.\ref{fig:fig2}(b) \cite{mishra2010fluctuations}. 

 {\em Steady state}:-
Now we explore the effect of birth and death on the steady state properties of the system. One  of the characteristics of immortal flock is the formation of ordered high density bands moving in the low density disordered background. These bands move as the sound wave travelling along the ordering direction  of the flock \cite{mishra2010fluctuations, chate2008collective, doi:10.1063/5.0086952, bhattacherjee2015topological}. 
In SM2, we show the animations of local density in the steady state for different $\lambda's$ = $0$, $0.0005$, $0.0008$, $0.001$ for system of size $L = 256$.
It can be observed  that the density fluctuations as well as the speed  of the moving bands decreases with increasing $\lambda$. To further quantify the motion of travelling density pattern for different $\lambda$, we define the one dimensional density profile $\rho_{\parallel}(x)$, 
by averaging the local density $\rho({\bf r}, t)$ over the direction transverse to the propagating
direction of the band i.e, $\rho_{\parallel}(x)= \frac{\sum_{j=1}^{L} \rho_{j, \perp}}{L}$;
here, $\rho_{\perp, \parallel}$ is the density along the transverse and longitudinal direction of the global orientation direction of the flock respectively. $x$ is chosen as the travelling direction (direction of ordering of the flock) of the bands.
In the Fig.\ref{fig:fig4} $(a-d)$, we show the plot of $\rho_{||} (x)$ for four different $\lambda$'s = $0$, $0.0005$, $0.0008$ and $0.001$ respectively. Different colors in each plot represents the $\rho_{\parallel}(x)$ at four different times at the interval of $t=400$. We find that for  $\lambda=0$, the pattern of $\rho_{\parallel} (x)$ shifts towards the direction of propagating band (shown by the arrow). The shift is proportional to the speed of the bands. As we introduce $\lambda$, and on increasing it, the shift  as well as the amplitude of fluctuation reduces
and for  $\lambda >0.0005$, the band fades away and density 
approaches to the mean value $\rho_{0}=0.75$ at late times.
We also calculated the sound speed using the linearised hydrodynamic equations of motion for the two fields (${\bf v}$ and $\rho$). The SM1, contains the details of the calculation. We find that sound speed very much depends on the birth and death rate and decreases with increasing $\lambda$ as given in Eq.14 (\ref{appendix}). Also interestingly we find that the sound speed is purely non-dispersive for IF ($\lambda=0$) and becomes dispersive for finite $\lambda$ Eq.14 (\ref{appendix}) (although the dependence is quadratic is $\lambda$). In \ref{appendix} we show the plot of $\rho_{\parallel}(x)$ for three different system sizes $K=175, 200, 256$ for $\lambda=0.0005$. Clearly the speed of the band depends on $K$ and increases on increasing $K$.

Further we also calculate the effect of $\lambda$ on the number fluctuation $\Delta N$ of particle density and  compared it with the density fluctuation calculated from the 
linearised approximation. The details of the linear calculation using coarse-grained hydrodynamic 
equations is given in \ref{appendix}.  
In Fig.\ref{fig:fig5}(a) we show the plot of $\frac{\Delta N}{\sqrt{N}}$ vs. $N$ for different $\lambda$ values. We expect  the $\frac{\Delta N}{\sqrt{N}} \simeq  N^{\alpha}$, where the exponent $\alpha > 0$ for density fluctuation larger than the corresponding equilibrium systems \cite{chaikin1995principles}.
We find that for $\lambda \le 0.05$, density fluctuation is larger than the equilibrium and for higher $\lambda >0.05$, it is suppressed. Further in Fig  \ref{fig:fig5}(b) we plot $\beta(N, \lambda)$ {\em vs.} $N$, where $\beta(N, \lambda)$ is obtained by $\beta(N, \lambda)= \frac{d\ln{\frac{\Delta N}{\sqrt{N}}}}{d \ln{N}}$. For small $N$, $\beta(N, \lambda)$   increases with $N$ and for large $N$ it approaches to value = $0.4$, for  $\lambda \le 0.05$ and zero for large $\lambda > 0.05$. In Fig. \ref{fig:fig5}(c) we show the plot of $\beta(\lambda)$ for different $\lambda$;
where $\beta(\lambda)$ is obtained from the large $N$ limit of $\beta(N, \lambda)$ from the previous plot. The data (black) shows the calculation from the numerical results and red data (lines) is the prediction from
the linearised hydrodynamics (\ref{appendix} Eq.21). We find that for small $\lambda$, linearised theory under predicts the behaviour of $\beta(\lambda)$, whereas for large $\lambda$  it approaches to zero and numerical result matches with theoretical prediction.  
    \section{Summary}
In this letter, we have modeled the Malthusian flock with birth and death rate in two-dimensions using coarse-grained hydrodynamic equations of motion for local velocity and density. Unlike the previous study of \cite{toner2012birth} for high density, our study is focused on the finite density of particles, such that density field cannot be ignored. Starting from the random homogeneous  state; the ordering kinetics of the system towards the ordered state is studied using numerical integration of the coupled non-linear partial differential equations of motion for the two fields. We find that the ordering kinetics of the velocity field remains unaffected due to birth and death rate, but density shows strong dependence. For IF and small birth and death rate the asymptotic growth law for density is $5/6$ and shows a crossover to diffusive growth law of $1/3$ for large birth and death rate. The approach to asymptotic growth becomes faster as we increase the birth and death rate. We also characterised the steady state properties of the flock using both numerical integration of full non-linear equations and linearised hydrodynamic about the ordered state and find that the presence of finite birth and death rate suppresses the speed of sound wave and large density fluctuations present in the immortal flock \cite{toner2012birth}. \\
Our study shows the effect of birth and death rate on the ordering kinetics and steady state properties of polar self-propelled particles for the case when density fluctuations are taken in account. Hence it can be used to understand the effect of birth and death in various biophysical systems with moderate densities \cite{voituriez2005spontaneous,kruse2005generic,mather2010streaming,karig2018stochastic,cates2010arrested}. Our finding can be tested for similar experiments\cite{dell2018growing} in the laboratory.  
\section{Acknowledgments}
P.J. gratefully acknowledge the DST INSPIRE fellowship  for
funding this project. The support and the resources provided by PARAM Shivay Facility under the National Supercomputing Mission, Government of India at the Indian Institute of Technology, Varanasi are gratefully acknowledged by all authors. S.M. thanks DST-SERB India, ECR/2017/000659 and CRG/2021/006945 for financial support. P.J. and S.M. also thank the Centre for Computing and Information Services at IIT (BHU), Varanasi.
\section{appendix}\label{appendix}
  \subsection{Ordering kinetics}\label{Ordering kinetics}
 \begin{figure} [h]
      \centering
      \includegraphics[scale=0.32]{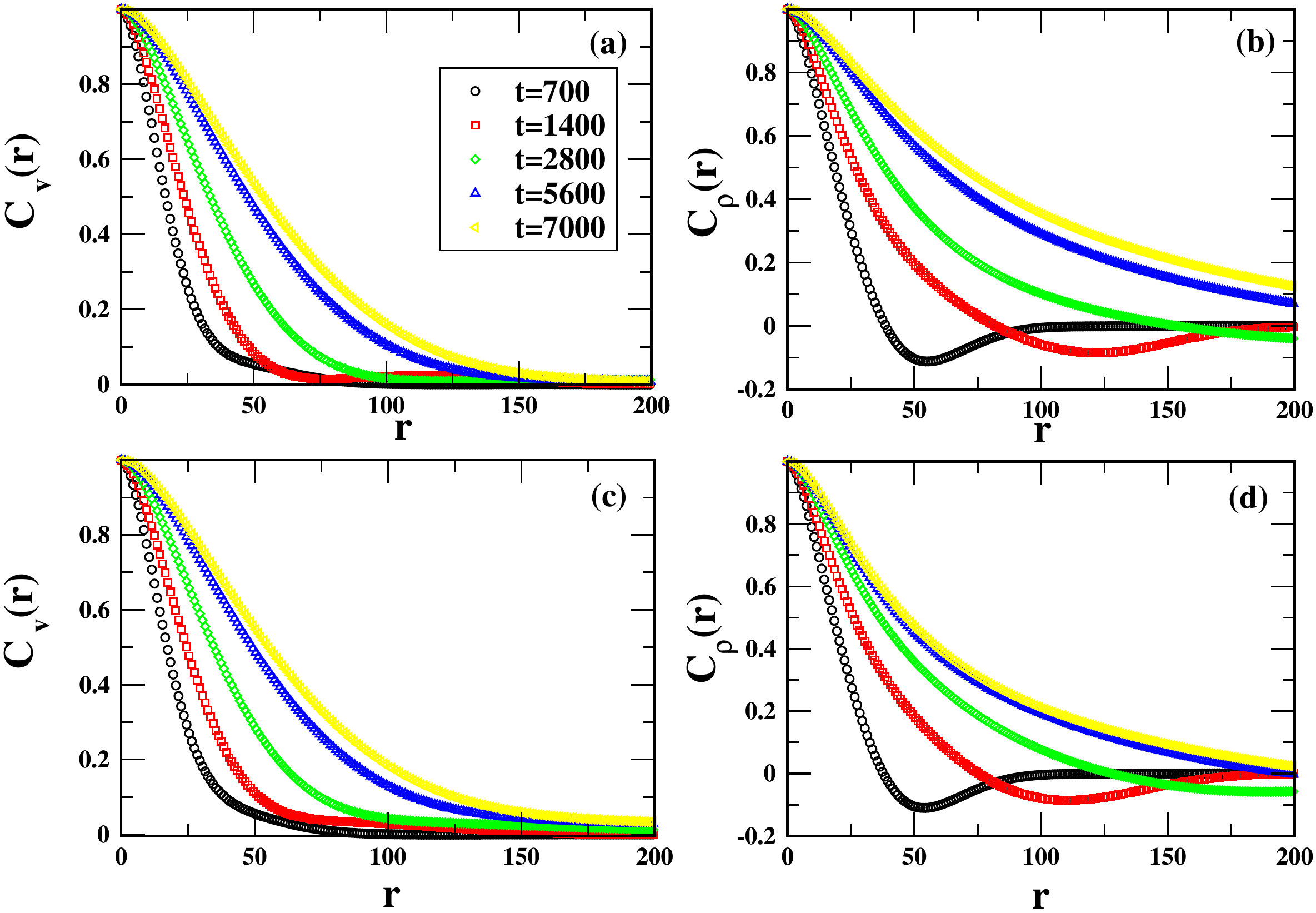}
      \caption{(color online) The plot of two-point correlation functions $C_{v, \rho}(r)$ {\em vs.} distance $r$ for velocity  and density fields.  $(a-b)$ is for the IF ($\lambda=0$) and $(c-d)$ is for the MF with $\lambda=0.01$. Different colors are for different times as shown in the legend.}
      \label{fig:figsm1}
  \end{figure}
 \subsection{Steady state}\label{Steady state}
\begin{figure}[h]
\centering
\includegraphics[scale=0.32]{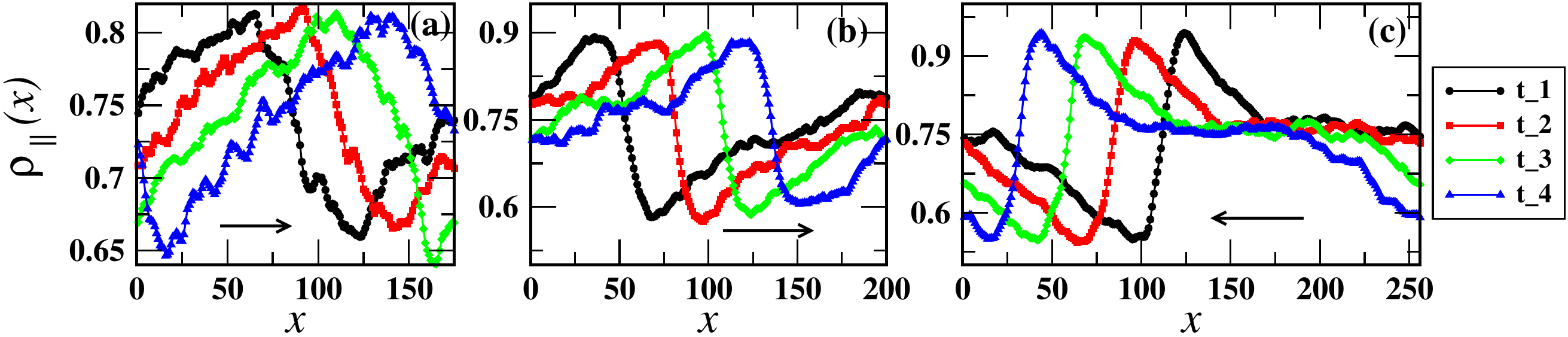}
\caption{(color online) The plot of $\rho_{||}(x)$  {\em vs.} $x$ for $\lambda=0.0005$ for three different system sizes $(a-c)$; $K=175$, $200$ and $256$ respectively for four different times $t_1 < t_2 < t_3 < t_4$ in the steady state at the time interval of $t=400$.  The arrows in each figure represent the direction of moving flock.}
      \label{fig:figsm2}
  \end{figure}
\subsection{Linearized approximation\label{Linearized approximation}}
In this section we focus on the linearised hydrodynamics of the system about the broken symmetry state. The steady state solution of Eq.1 and Eq.2 in the main text is a homogeneous ordered state with density $\rho=\rho_{0}$ and order parameter $ {\bf v}= v{_0}\widehat{x}$, where $v_0 = \sqrt{\frac{\alpha}{\beta}}$ and $\alpha=\alpha_0(1-\rho_0/\rho_c)$. Here we have chosen $\hat{x}$ as the  direction of broken symmetry. 
We  add small fluctuations to the uniformly ordered velocity and homogeneous density, 
 and write the velocity field as ${\bf v}$=$ v{_0}\widehat{x}+\delta{\bf v}$. The fluctuation in ${\bf v}$, $\delta {\bf v} = (v_{||}\widehat{x}, v_{\perp} \widehat{y})$, where $\hat{y}$ is direction perpendicular to the ordering direction.
 Now we write the equation of motion for the fluctuation $\delta {v_{||}}$ to the linear order in small fluctuations $(\delta \rho,  {v_{||}}, {v_{\perp}})$
\begin{equation}\label{Eq:delv}
    \partial{_t} v{_\parallel}= -\sigma{_1}\partial{_\parallel}\delta \rho-2\alpha  v{_\parallel}+ \text{irrelevant terms}
\end{equation}
In the steady state, we can solve the right hand side of Eq. \ref{Eq:delv} and get the following solution for $ v_{||}$

\begin{equation}
   v_{\parallel}=-D_{2}\partial_{||}\delta\rho
   \label{Eq:delv2}
\end{equation}
where,
$D_{2} = \sigma_{1}/2\alpha$ .
Inserting Eq.\ref{Eq:delv2} in the linearised equations of motion for ${v}_{\perp}$ and $\delta\rho$ and neglecting the  irrelevent terms,

\begin{equation}
\partial_{t} v_{\perp} + \gamma \partial_{||} v_{\perp} = - \sigma{_1} \partial_{\perp} \delta \rho +  D_{T}\nabla^2 v_{\perp} + f_{\perp}
\label{Eq:velperp}
\end{equation} 

and density equation in Eq.1(main text), 
\begin{dmath}
 \partial_{t} \delta \rho + v_{0} \partial_{||} \delta \rho - D_{2} \rho_{0} \partial_{||}^2 \delta \rho -D_{\rho}\nabla^{2} \delta \rho+ \rho_{0} \partial_{\perp} v_{\perp} = -\lambda \rho_{0} \delta \rho
\label{Eq:delrho}
\end{dmath}                  
where $D_{2}, D_{\rho},D_{B}$ and $D_{T}$ are the diffusion constants, $\gamma=\lambda_{1}v_{0}$. For the given system parameters, they are all constants. To further
study the solution of above linearised equations we define a vector ${\bf A}({\bf r}, t) = (v_{\perp}, \delta \rho)$. Eq. \ref{Eq:velperp} and Eq.\ref{Eq:delrho}, we write the equations in Fourier space by defining
\begin{equation*}
{\bf A}({\bf r}, t) \simeq  \int{{\bf A}({\bf q},\omega) \exp(-i\omega t+ i{\bf q}.{\bf r}) d{\bf q} d \omega}
  \end{equation*}
  Now we can write the linear equations for $v_{\perp}({\bf q}, \omega)$ and $\delta \rho({\bf q}, \omega)$ 
\begin{equation}
   [-i(\omega-\gamma q{_\parallel})+\Gamma{_L}]v{_\perp}+i\sigma{_1}q{_\perp}\delta\rho=f{_\perp} 
\label{Eq:vf}
\end{equation}
and 
\begin{dmath}\label{Eq:rhof}
[-i(\omega-v{_0} q{_\parallel})+\Gamma{_\rho}+ \lambda\rho_{0}]\delta\rho+i\rho{_0}q{_\perp}v{_\perp}=0
\end{dmath}

where ${\bf q}$ = $(q_{\parallel}, q_{\perp})$ =  $q(\cos (\theta), \sin(\theta))$, 
where $\theta$ is the angle from the direction of broken symmetry. The Fourier transform of random force $f_{\perp}({\bf q},\omega)$ has properties $<f_{\perp}({\bf q}, \omega)> =0 $ and variance $<f_{\perp}({\bf q}, \omega) f_{\perp}({\bf q}', \omega')> = \Delta \delta({\bf q}+{\bf q}')\delta(\omega+\omega') $; The two damping  $\Gamma_{L,\rho}({\bf q})$ defined as $\Gamma{_L}({\bf q})= D{_T}(q_{\parallel}^{2}+q_{\perp}^{2})$ and  $\Gamma_{\rho}({\bf q})=D{_2}\rho_{0} q_{\parallel}^{2}+D_{\rho}(q_{\parallel}^{2}+q_{\perp}^{2})$. The linear equations Eq.\ref{Eq:vf} and Eq.\ref{Eq:rhof} for $v_{\perp}$ and $\delta\rho$ can be solved using following matrix; 
\begin{widetext}
\[
  \begin{bmatrix}
    -i(\omega-\gamma q_{\parallel})+\Gamma_{L}  &       i\sigma_{1} q_{\perp} \\
    i q_{\perp}\rho_{0}                         &  -i(\omega-v_{0} q_{\parallel})+\Gamma_{\rho}+\lambda \rho_{0}          
     \end{bmatrix}
  \begin{bmatrix}
    v_{\perp} \\ \delta\rho
  \end{bmatrix}
  = \begin{bmatrix}
    f_{L}  \\ 0
      
  \end{bmatrix}
 \]
 The solutions for $v_{\perp}$ and $\delta\rho$ is
 \[
 \begin{bmatrix}
  v_{\perp} \\ \delta\rho
 \end{bmatrix}
 =
 [M]^{-1}
 \begin{bmatrix}
  f_{T}  \\ 0
 \end{bmatrix}
 \]
 where the matrix $[M]$ is,

\[
\begin{bmatrix}
 M
 \end{bmatrix}
  =
 \begin{bmatrix}
  -i(\omega-\gamma q_{\parallel})+\Gamma_{L}  &       i\sigma_{1} q_{\perp} \\
    iq_{\perp}\rho_{0}                         &  -i(\omega-v_{0} q_{\parallel})+\Gamma_{\rho}+\lambda \rho_{0}
 \end{bmatrix}
 \]
 
and $[M]^{-1}$ is   
 \[
 \begin{bmatrix}
 M^{-1}
 \end{bmatrix}
 = 
 \frac{1}{\det[M]}
 \begin{bmatrix}
  -i(\omega-v_{0} q_{\parallel})+\Gamma_{\rho}+\lambda \rho_{0} &    - i\sigma_{1}q_{\perp}   \\
       - i q_{\perp}\rho_{0}                    &  -i(\omega-\gamma q_{\parallel})+\Gamma_{L}
 \end{bmatrix}
 \]
\end{widetext}
 where
 \begin{dmath}
  det[M]= -\omega^{2}+\omega(v_{0}q_{\parallel}+\gamma q_{\parallel}-i\Gamma_{\rho}-i\Gamma_{L}-i \lambda \rho_{0})-v_{0}\gamma q_{\parallel}^{2}+i q_{\parallel}(\gamma\Gamma_{\rho}+v_{0}\Gamma_{L}) +\Gamma_{\rho}\Gamma_{L}+\rho_{0}\sigma_{1}q_{\perp}^{2}+\Gamma_{L} \lambda \rho_{0}+ i \lambda \rho_{0}\gamma q_{\parallel}
\end{dmath}
The quadratic equation for the $\omega$;
\begin{dmath}
    \omega^2-\omega[q_{\parallel}(v_{0}+\gamma)- i(\Gamma_{\rho}+\Gamma_{L}+\lambda \rho_{0})]-i q_{\parallel}(\gamma\Gamma_{\rho}+v_{0}\Gamma_{L})+v_{0}\gamma q_{\parallel}^2-\rho_{0}\sigma_{1}q_{\perp}^2
-\Gamma_{L}\lambda\rho_{0}-i \lambda\rho_{0}\gamma q_{\parallel}=0
\end{dmath}

The real part two solutions of $\omega_{\pm}$ are given by;  
\begin{dmath}
    \omega_{\pm}= c_{\pm}(\theta_{{\bf q}})q-\frac{\Delta \Gamma \lambda\rho_{0} q}{\sqrt{(\frac{\gamma-v_{0}}{2})^2\cos^2{\theta}+c_{0}^2\sin{\theta}}} -\frac{(\lambda\rho_{0})^2}{2q {\sqrt{(\frac{\gamma-v_{0}}{2})^2\cos^2{\theta}+c_{0}^2\sin{\theta}}}}
\label{Eq:eqomegapm}
\end{dmath}
where
\begin{equation*}
  c_{\pm}(\theta_{{\bf q}})=\frac{\gamma+v_{0}}{2} \cos(\theta _{{\bf q}})\pm c_{2}(\theta _{{\bf q}})
  \label{Eq:sp}
  \end{equation*}
  \begin{equation*}
  c_{2}(\theta_{{\bf q}})=\sqrt{\frac{1}{4}(\gamma-v_{0})^{2}\cos^{2}(\theta_{{\bf q}})+c_{0}^{2}\sin^{2}(\theta_{{\bf q}})}
  \end{equation*}
  \begin{equation*}
  c_{0}=\sqrt{\sigma_{1}\rho_{0}}
  \end{equation*}
and $\Delta \Gamma = \widehat{\Gamma_{\rho}} - \widehat{\Gamma_{L}}$,
$\widehat{\Gamma_{L}}=D_{T}$,
$\widehat{\Gamma_{\rho}}= D_{2}\rho_{0}\cos{\theta}+D_{\rho}$.We can calculate the speed of sound wave $C(q, \lambda,\theta)= \frac{d \omega_{+}}{d t}$ from Eq. \ref{Eq:eqomegapm} as 

\begin{dmath}
    C(q, \lambda,\theta)= \frac{\gamma+v_{0}}{2} \cos(\theta _{{\bf q}})+\sqrt{\frac{1}{4}(\gamma-v_{0})^{2}\cos^{2}(\theta_{{\bf q}})+c_{0}^{2}\sin^{2}(\theta_{{\bf q}})}-\frac{\Delta \Gamma \lambda\rho_{0}}{\sqrt{\frac{1}{4}(\gamma-v_{0})^{2}\cos^{2}(\theta_{{\bf q}})+c_{0}^{2}\sin^{2}(\theta_{{\bf q}})}}+\frac{( \lambda\rho_{0})^2}{q^2 \sqrt{\frac{1}{4}(\gamma-v_{0})^{2}\cos^{2}(\theta_{{\bf q}})+c_{0}^{2}\sin^{2}(\theta_{{\bf q}})}}
\end{dmath}
For the case $\theta=0$ , the speed of sound wave from Eq.\ref{Eq:eqomegapm} as
\begin{equation}
   C(q, \lambda) = \frac{d\omega_{+}}{dq}= v_0 -\frac{2\Delta \Gamma \lambda\rho_{0}}{(\gamma-v_{0})}+\frac{4( \lambda\rho_{0})^2}{q^2({\gamma-v_{0}})^2} 
\end{equation}
Further we go back to the solution for $v_{\perp}$ and $\delta \rho$
  \begin{equation}\label{Eq:vs}
       v_{\perp}({\bf q}, \omega)=G_{LL}({\bf q}, \omega) f_{\perp}({\bf q}, \omega)
  \end{equation}
  \begin{equation}
      \delta \rho({\bf q}, \omega) = G_{\rho L} ({\bf q}, \omega)f_{\perp}({\bf q}, \omega)
  \end{equation}
  where the propagators $G_{LL}$ and $G_{\rho L}$ are given by;
  \begin{widetext}

  \begin{equation}
      G_{LL}({\bf q}, \omega)=\frac{(-i(\omega-v_{0}q_{\parallel})+\Gamma_{\rho}({\bf q})+\lambda \rho_0)}{(\omega-c_{+}(\theta_{\bf q})q)(\omega-c_{-}(\theta_{\bf q})q)+i \omega(\Gamma_{L}({\bf q})+\Gamma_{\rho}({\bf q})+\lambda\rho_{0})-i q\cos{\theta}(\gamma \Gamma_{L}({\bf q})+v_{0} \Gamma_{\rho}({\bf q})+2\lambda\rho_{0}(\gamma+v_{0}))}
  \end{equation}
   \begin{equation}
      G_{\rho L}({\bf q}, \omega) = \frac{-i \sigma_{1}q_{\perp}}{(\omega-c_{+}(\theta_{\bf q})q)(\omega-c_{-}(\theta_{\bf q})q)+i \omega(\Gamma_{L}({\bf q})+\Gamma_{\rho}({\bf q})+\lambda\rho_{0})-i q\cos{\theta}(\gamma \Gamma_{L}({\bf q})+v_{0} \Gamma_{\rho}({\bf q})+2\lambda\rho_{0}(\gamma+v_{0}))}
  \end{equation}
  Hence, the two-point density-density correlation function $C_{\rho}({\bf q}, \omega)$ with become;
  
\begin{equation}
C_{\rho}(q,\omega)=\frac{\Delta\rho_{0}^{2}q_{\perp}^{2} }{{(\omega-c_{+}(\theta_{\bf q})q)^{2}(\omega-c_{-}(\theta_{\bf q})q)^{2}+ (\omega(\Gamma_{L}({\bf q})+\Gamma_{\rho}({\bf q})+\lambda\rho_{0})- q\cos{\theta}(\gamma \Gamma_{L}({\bf q})+v_{0} \Gamma_{\rho}({\bf q}))+2\lambda\rho_{0}(\gamma+v_{0}))^{2}}}
\end{equation}
\end{widetext}
 The $C_{\rho}$  has two sharp peaks at $\omega$= $c_{\pm}$q with width $\mathcal{O}(q^{2})$.
  The width of these peaks in the limit $q \to 0$, is very small compared to its displacement from the origin ($\mathcal{O}(\omega)$). The expression for the structure factor $S(q) = C_{\rho }(q) =  \int_{-\infty}^{+\infty} C_{\rho }({\bf q},\omega) \,\frac{d\omega}{2\pi}$ is
\begin{dmath}
S(q) = \frac{\Delta_{0}\rho_{0}^{2}q_{\perp}^{2}+ c_{+}^{2}q^{2}\lambda^{2} \rho_{0}^{2}}{[c_{+}q(\Gamma_{L}({\bf q})+\Gamma_{\rho}({\bf q})+\lambda\rho_{0})]^{2}} \times [1+ \frac{\Gamma_{\rho}({\bf q})}{\Gamma_{L}({\bf q})}+\frac{2\lambda\rho_{0}(\gamma+v_{0}+0.5)}{\Gamma_{L}({\bf q})}]
\end{dmath}
  Here we have neglected the higher order term of $\mathcal{O}(q^{4})$. We write the expression for small $\lambda$ such that  we can ignore the term of order $\mathcal{O}(\lambda ^2)$.
Writing the different values of constants from the numerical simulation we get,
\begin{equation*}
    S(q)=\frac{A[q^2+B\lambda]}{[Cq^{2}+D\lambda]^{2}}
    \label{Eq:sq}
\end{equation*}
where $A=0.056$, $B=2.25$, $C=0.86$, $D=0.75$.
  We can now calculate the  number fluctuation $\Delta N$  in a volume $V$ as $\sqrt{S(q\to 0)V}$. We get the expression for $\Delta N$  as
  \begin{equation}
     \Delta N= N \frac{\sqrt{ (A(1+BN \lambda))}}{C+D N\lambda} 
     \label{Eq:deln}
  \end{equation}

\end{document}